\documentclass[twoside,fleqn]{ActaStyle}
\usepackage{times,cite}

\usepackage{lscape}             
\usepackage{graphicx,epsfig}    
\usepackage[tbtags]{amsmath}    
\usepackage{hhline}
\usepackage{dcolumn}

\def\authorheading{B.~Borasoy and R.~Ni{\ss}ler}

\def\shorttitle{Decays of $\eta$ and  $\eta'$}

\renewcommand{\pagerange}[2]{%
\newcount{\prvastrana}\prvastrana=#1
\newcount{\poslednastrana}\poslednastrana=#2
\setcounter{page}{#1}}

\renewcommand{\publ}{\begin{minipage}{0.9\textwidth}
Talk presented at the Eta'05 Workshop on production and decay
of $\eta$ and $\eta'$ mesons,
September 2005, Cracow, Poland.
\end{minipage}
}

\begin{document}

\pagerange{1}{7}

\title{%
Decays of $\mbox{\boldmath$\eta$}$ and $\mbox{\boldmath$\eta'$}$
within a chiral unitary approach}

\author{
B.~Borasoy\email{borasoy@itkp.uni-bonn.de} and
R.~Ni{\ss}ler\email{rnissler@itkp.uni-bonn.de}
}
{
Helmholtz-Institut f\"ur Strahlen- und Kernphysik (Theorie) \\
Universit\"at Bonn,
Nu{\ss}allee 14-16, D-53115 Bonn, Germany }


\abstract{%
Various decays of $\eta$ and  $\eta'$
are investigated within the framework of U(3) chiral effective field theory
in combination with a relativistic coupled-channels approach.
Final state interactions are included by deriving $s$- and $p$-wave interaction
kernels for meson-meson scattering from the chiral effective Lagrangian and
iterating them in a Bethe-Salpeter equation.
Very good agreement with experimental data is achieved.}


\section{Introduction}
\label{sec:intr} \setcounter{section}{1}\setcounter{equation}{0}

Decays of $\eta$ and $\eta'$ offer a possibility to study symmetries
and symmetry breaking patterns in strong interactions.
The isospin violating decays $\eta, \eta' \to 3 \pi$, e.g., can only occur
due to an isospin breaking quark mass difference $m_u-m_d$ or electromagnetic
effects. While for most processes isospin violation of the strong interactions is
masked by electromagnetic effects, these corrections are expected to be small for the
three pion decays of $\eta$ and $\eta'$ (Sutherland's theorem) \cite{Sutherland:1966mi}.
Neglecting electromagnetic corrections the decay amplitude is directly proportional to 
$m_u-m_d$. The decays $\eta, \eta' \rightarrow \gamma \gamma$ and
$\eta, \eta' \to \pi^+ \pi^- \gamma$, on the other hand,
are phenomenological manifestations of the chiral anomaly of QCD and
can provide important information on the chiral symmetry of the strong 
interactions.

Moreover, the $\eta$-$\eta'$ system offers a testing ground for 
the role of both spontaneous and explicit chiral symmetry breaking,
the latter one induced by the light quark masses.
In the absence of $\eta$-$\eta'$ mixing, $\eta$ would be the pure member $\eta_8$ of the octet
of Goldstone bosons which arise due to spontaneous breakdown of chiral symmetry.

Reactions involving the $\eta'$
might also provide insight into gluonic effects through the axial U(1) anomaly of QCD.
The divergence of the singlet axial-vector current acquires an additional term
with the gluonic field strength tensor that remains
in the chiral limit of vanishing light quark masses.
This term prevents the pseudoscalar
singlet $\eta_0$ from being a Goldstone boson which is phenomenologically
manifested in its relatively large mass, $m_{\eta'} = 958$ MeV.

An appropriate theoretical framework to investigate low-energy hadronic physics is
provided by chiral perturbation theory (ChPT) \cite{GL}, the effective field theory 
of QCD. In ChPT Green's functions are expanded perturbatively in powers of Goldstone boson
masses and small three-momenta. 
However, final state interactions, e.g., in $\eta \to 3 \pi$ 
have been shown to be substantial both in a complete one-loop calculation
in SU(3) ChPT \cite{GL2} and using extended Khuri-Treiman equations \cite{Kambor}.
In $\eta'$ decays final state interactions are expected to be even more important
due to larger phase space and the presence of nearby resonances.
It is claimed, e.g., that the exchange of the scalar resonance $a_0(980)$ dominates
the decays $\eta' \to \eta \pi \pi$ \cite{Fariborz} which has been
confirmed both in a full one-loop calculation utilizing infrared regularization \cite{BB2}
and in a chiral unitary approach \cite{BB1}.

Also for the decays $\eta, \eta' \rightarrow \gamma^* \gamma^{(*)}$ and
$\eta' \to \pi^+ \pi^- \gamma$ contributions from vector meson exchange
will dominate the amplitude and unitarity effects should be implemented via final
state interactions. 
The Dalitz decays $\eta, \eta' \rightarrow \gamma^* \gamma$ probe the charge
distribution of $\eta$ and $\eta'$, while the double Dalitz decays
$\eta, \eta' \rightarrow \gamma^* \gamma^*$ may indicate whether double vector dominance is
realized in nature.
In particular the $\eta'$ decays are clearly beyond the perturbative framework of chiral
perturbation theory (ChPT) and require utilization of non-perturbative tools
which match onto the results from ChPT.

In a series of papers \cite{BB1, BN1, BN2, BN3} we have developed a theoretical framework
which describes these $\eta, \eta'$ decays in a uniform manner. The approach is based
on U(3) chiral effective field theory
in combination with a relativistic coupled-channels approach.
Final state interactions are included by deriving $s$- and $p$-wave interaction
kernels for meson-meson scattering from the chiral effective Lagrangian and
iterating them in a Bethe-Salpeter equation (BSE). The illustration of the underlying idea
of this method as well as a summary of the results for the hadronic decays
is the subject of this presentation.

\section{Sketch of the approach}
\label{sec:sketch}

\begin{figure}[hbt]
\centering\includegraphics[width=0.3\textwidth]{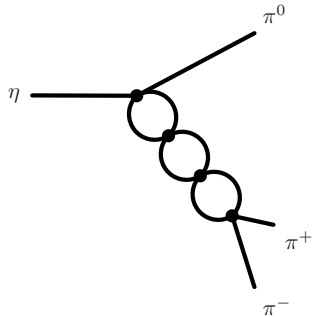}
\caption{Shown is a possible contribution to final state interactions in the decay 
$\eta \to \pi^+ \pi^- \pi^0$.}
\label{fig:Method}
\end{figure}

Let us start with the hadronic decays $\eta, \eta' \to 3\pi$ and $\eta' \to \eta \pi \pi$ \cite{BB1, BN3}.
The underlying idea of the approach is that the initial particle, 
i.e.\ the $\eta$ or $\eta'$, decays into three 
mesons and that two out of these rescatter (elastically or inelastically) an arbitrary number of times, see 
Fig.~\ref{fig:Method} for illustration. All occurring vertices are derived 
from the effective Lagrangian and are 
thus constrained by chiral symmetry. 
Interactions of the third meson with the pair of rescattering mesons are neglected which turns out to be a 
good approximation, particularly for the decays $\eta \to 3\pi$ and $\eta' \to \eta \pi \pi$.

Such an infinite meson-meson rescattering process can be 
generated by application of the Bethe-Salpeter equation.
To this aim, the partial wave interaction kernels for meson meson scattering, $A_l$, are derived from
the chiral effective Lagrangian and iterated in a Bethe-Salpeter equation
which generates the propagator for two interacting particles in a covariant
fashion.
For each partial wave $l$ the matrix-valued solution $T_l$ of the BSE with coupled channels 
and on-shell interaction kernels is given by
\begin{equation}
T^{-1}_l = A^{-1}_l + G \ ,
\end{equation}
where the diagonal matrix G collects the scalar loop integrals of the different two-meson channels.
The solution of the BSE satisfies exact unitarity for two-particle scattering
and generates resonances dynamically by an infinite string of meson-meson rescattering processes,
see also \cite{OO}.
In the investigations discussed here we have restricted ourselves to $s$- and $p$-wave
interaction kernels.

\begin{figure}[hbt]
\centering
\includegraphics[width=0.3\textwidth]{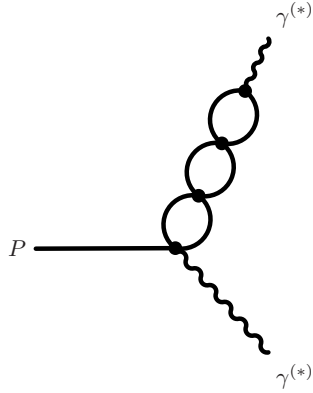}
\caption{Sample rescattering process in the decay $P \to \gamma^{(*)} \gamma^{(*)}$.}
\label{fig2gam}
\end{figure}

This approach can be extended to the anomalous decays  
$\eta, \eta' \rightarrow \gamma^{(*)} \gamma^{(*)}$ \cite{BN1} and
$\eta, \eta' \to \pi^+ \pi^- \gamma$ \cite{BN2} in a straightforward manner.
In $\eta, \eta' \rightarrow \gamma^{(*)} \gamma^{(*)}$, e.g., the incoming pseudoscalar meson $P$
can directly decay via a vertex of either the Wess-Zumino-Witten Lagrangian, which describes the
chiral anomaly,
or the unnatural-parity Lagrangian
into one of the following three channels: two photons, a photon and two pseudoscalar mesons, 
or four pseudoscalar mesons.
Pairs of mesons can then rescatter an arbitrary number of times
before they eventually couple to a photon,
see Fig.~\ref{fig2gam} for illustration.
In order to implement the non-perturbative summation of loop graphs 
covered by the BSE in the decay processes
under consideration, we treat the BSE $T$-matrix as an effective vertex for meson-meson scattering. 
The anomalous decays into $\pi^+ \pi^- \gamma$ are treated in the same way \cite{BN2}.

\section{Results}
\label{sec:results}

We now turn to the discussion of the numerical results, where for brevity we restrict ourselves
to the hadronic decays\cite{BN3}. Results for the anomalous decays can be found in \cite{BN1,BN2}. 
The results of our calculation for the hadronic decays are obtained 
from a combined 
analysis of the decay widths, branching ratios, and slope parameters of the considered decays as well as 
phase shifts in meson-meson scattering. The widths of $\eta \to 3 \pi$ and 
$\eta' \to \eta \pi \pi$ have been 
measured roughly at the 10\,\% precision level,
while for $\eta' \to 3 \pi^0$ the experimental uncertainty is considerably larger and only an 
upper limit exists for $\Gamma(\eta' \to \pi^+ \pi^- \pi^0)$ \cite{pdg}. 
Moreover, some of these decay widths are 
constrained by the well-measured branching ratios
\begin{equation} \label{eq:defBR}
r_1 = \frac{\Gamma(\eta \to 3\pi^0)}{\Gamma(\eta \to \pi^+ \pi^- \pi^0)} \,, \qquad
r_2 = \frac{\Gamma(\eta' \to 3\pi^0)}{\Gamma(\eta' \to \eta \pi^0 \pi^0)} \,.
\end{equation}

The squared absolute value of the decay amplitude, 
$|\mathcal{A}(x,y)|^2$, is expanded for $\eta' \to \eta \pi \pi$ and the charged decay modes
of $\eta, \eta' \to 3\pi$ in terms of the usual Dalitz variables $x,y,$ as
\begin{equation} \label{eq:DalC}
|\mathcal{A}(x,y)|^2  = |N|^2 \big[ 1 + a y + b y^2 + c x^2 + d y^3 + \cdots \big]\, ,
\end{equation}
while for the decays into three identical particles Bose symmetry dictates the form
\begin{equation} \label{eq:Dal0}
|\mathcal{A}(x,y)|^2  = |N'|^2 \big[ 1 + g( y^2 + x^2) + \cdots \big] \,.
\end{equation}
For the Dalitz plot parameters $a$, $b$, $c$ of $\eta \to \pi^+ \pi^- \pi^0$ the experimental situation 
is not without controversy. We employ the numbers of \cite{Abele}, since it is the most recent published 
measurement and the results appear to be consistent with the bulk of the other experiments listed by the 
Particle Data Group \cite{pdg}. They differ somewhat from 
the new preliminary results of the KLOE Collaboration \cite{KLOE} that has found a non-zero value for the 
third order parameter $d$ which was not included in previous experimental parametrizations.
Very recently the Dalitz plot parameters 
of $\eta' \to \eta \pi^+ \pi^-$ have been determined with high statistics by the VES experiment \cite{VES}.
In this presentation, we will discuss the fits obtained
from the experimental Dalitz parameters provided by the Particle Data Group
\cite{pdg}.
Note that the slope parameters of $\eta' \to 3 \pi$ have not yet been determined experimentally, but such a 
measurement is intended at WASA@COSY \cite{WASA}.

{From} the unitarized partial-wave $T$-matrix one may also derive the phase shifts 
in meson-meson scattering. Hence, our approach is further constrained by the experimental phase shifts for 
$\pi \pi \to \pi \pi, K \bar{K}$ scattering. 

The coupled-channels framework entails several chiral couplings of the  
effective Lagrangian up to fourth order which must be fit to experimental data.
By fitting to all available (published) data sets of the investigated hadronic $\eta, \eta'$ decays
and phase shifts 
an overall $\chi^2$ function is calculated.
We observe four different classes of fits which are all in very good agreement with the 
currently available (published) data on hadronic decays, but differ in the description of the decays
$\eta' \to 3\pi$ where experimental constraints are scarce.
The errors which we specify in the 
following for all parameters and observables reflect the deviations 
which arise when we allow for $\chi^2$ values which are at most 15\,\% larger than the minimum value.
Although this choice is somewhat arbitrary, it illustrates how variation 
of the $\chi^2$ function in parameter space affects the results. 
The results for $\eta \to 3 \pi$ are shown in Table~\ref{tab:eta3pi}.

\begin{table}[hbt]
\centering
\begin{tabular}{|c|D{+}{\,\pm\,}{5,5}|D{+}{\,\pm\,}{4,4}|D{+}{\,\pm\,}{4,4}|D{+}{\,\pm\,}{6,6}|}
\hhline{----}
& \multicolumn{1}{c|}{$\Gamma_{\eta \to 3\pi^0}$ (eV)} & 
\multicolumn{1}{c|}{$\Gamma_{\eta \to \pi^+ \pi^- \pi^0}$ (eV)} &
\multicolumn{1}{c|}{$r_1$} \\
\hhline{----}
theo. &
  422    +    13    &
  290    +     8    &
  1.46   +   0.02   \\
\hhline{----}
exp. &
  419    +    27    &
  292    +    21    &
  1.44   +   0.04   \\
\hhline{====-}
& \multicolumn{1}{c|}{$a$} &
\multicolumn{1}{c|}{$b$} &
\multicolumn{1}{c|}{$c$} &
\multicolumn{1}{c|}{$g$} \\
\hline
theo. &
 -1.20   +   0.07   &
  0.28   +   0.05   &
  0.05   +   0.02   &
 -0.062  +   0.006  \\
\hline
exp. &
 -1.22   +   0.07   &
  0.22   +   0.11   &
  0.06   +   0.02   &
 -0.062  +   0.008  \\
\hline
\end{tabular}
\caption{Results for the partial decay widths of $\eta \to 3\pi$, the branching ratio $r_1$, and
         the Dalitz plot parameters compared to experimental data from \cite{pdg} and \cite{Abele}.}
\label{tab:eta3pi}
\end{table}

Only sparse experimental information exists on the decays of $\eta'$ into three pions. 
The experimental decay width of $\eta' \to 3 \pi^0$ is \cite{pdg}
\begin{equation}
\Gamma^{\textrm{(exp)}}(\eta' \to 3 \pi^0) = (315 \pm 78) \mbox{eV}
\end{equation}
which is nicely met within our approach:
\begin{equation}
\Gamma^{\textrm{(theo)}}(\eta' \to 3 \pi^0) = (330 \pm 33) \mbox{eV} \,.
\end{equation}
For the decay into $\pi^+ \pi^- \pi^0$ only a weak experimental upper limit exists, 
$\Gamma^{\textrm{(exp)}}(\eta' \to \pi^+ \pi^- \pi^0) < 10 \, \mbox{keV}$ \cite{pdg}.
As already mentioned above, we observe four different clusters of fits which
lead to equally good agreement with the available data but yield different
predictions for the $\eta' \to \pi^+ \pi^- \pi^0$ decay width and the unmeasured
$\eta' \to 3 \pi$ Dalitz parameters, see \cite{BN3} for details.
The Dalitz plot distributions of these decays and the decay width $\eta' \to \pi^+ \pi^- \pi^0$ 
pose therefore tight constraints for our approach and 
must be compared with future experiments at the WASA@COSY facility.

For $\Gamma_{\eta' \to \eta 2\pi}$ we obtain the results

\begin{equation}
r_{2}^{\textrm{(theo)}} = (71 \pm  7)\times 10^{-4}, \qquad 
r_{2}^{\textrm{(exp)}}  = (74 \pm 12)\times 10^{-4}  \quad \mbox{\cite{pdg}}.
\end{equation}

\begin{table}[hb]
\centering
\begin{tabular}{|c|r@{$\,\pm\,$}l@{\ }l|r@{$\,\pm\,$}l|r@{$\,\pm\,$}l|r@{$\,\pm\,$}l|}
\hline
& \multicolumn{3}{c|}{$\Gamma_{\eta' \to \eta \pi^+ \pi^-}$} &
\multicolumn{2}{c|}{$a$} &
\multicolumn{2}{c|}{$b$} &
\multicolumn{2}{c|}{$c$} \\
\hline
theo. &
$ 81   $ & $  4   $ & keV &
$-0.116$ & $ 0.024$ &
$ 0.000$ & $ 0.019$ &
$ 0.016$ & $ 0.035$ \\
\hline
exp. &
$ 89   $ & $ 11   $ & keV &
$-0.16 $ & $ 0.06 $ &
\multicolumn{2}{l|}{ } & 
\multicolumn{2}{l|}{ } \\
\hline
\end{tabular}
\caption{Results for the partial decay width of $\eta' \to \eta \pi^+ \pi^-$ and the Dalitz plot 
         parameters compared to experimental data from \cite{pdg}.}
\label{tab:9812}
\end{table}

\begin{table}[h]
\centering
\begin{tabular}{|c|r@{$\,\pm\,$}l@{\ }l|r@{$\,\pm\,$}l|r@{$\,\pm\,$}l|r@{$\,\pm\,$}l|}
\hline
& \multicolumn{3}{c|}{$\Gamma_{\eta' \to \eta \pi^0 \pi^0}$} &
\multicolumn{2}{c|}{$a$} &
\multicolumn{2}{c|}{$b$} &
\multicolumn{2}{c|}{$c$} \\
\hline
theo. &
$ 46   $ & $  3   $ & keV &
$-0.122$ & $ 0.025$ &
$ 0.003$ & $ 0.018$ &
$ 0.019$ & $ 0.039$ \\
\hline
exp. &
$ 42   $ & $  6   $ & keV &
$-0.116$ & $ 0.026$ &
$ 0.003$ & $ 0.017$ &
$ 0.00 $ & $ 0.03 $ \\
\hline
\end{tabular}
\caption{Results for the partial decay width of $\eta' \to \eta \pi^0 \pi^0$ and the Dalitz plot 
         parameters compared to experimental data from \cite{pdg}.}
\label{tab:9833}
\end{table}

In the isospin limit, $m_u = m_d$, the decay width $\Gamma(\eta' \to \eta \pi^+ \pi^-)$ would be exactly given by 
2 $\Gamma(\eta' \to \eta \pi^0 \pi^0)$, due to the symmetry factor for identical particles in the latter process. 
If, however, one is interested in the isospin-breaking contributions in the amplitude of $\eta' \to \eta \pi \pi$,
one ought to disentangle it from phase space effects which are caused by the different masses of charged and neutral
pions. With an isospin-symmetric decay amplitude, but physical masses in the phase space factors, we find a ratio
\begin{equation}
r_3 = \frac{\Gamma(\eta' \to \eta \pi^+ \pi^-)}{\Gamma(\eta' \to \eta \pi^0 \pi^0)} = 1.78 \pm 0.02 \,,
\end{equation}
which is smaller than 2 and compares to $r_3 = 1.77 \pm 0.02$
when isospin-breaking is taken into account in the amplitude.
(For comparison, if the amplitude is set constant and the physical pion masses are employed in the phase 
space integrals, the ratio is given by $r_3 = 1.77$.)
We may thus conclude that within our approach 
isospin-breaking corrections in the $\eta' \to \eta \pi \pi$ decay amplitude are tiny.
The branching ratio $r_3$ has not been measured directly. If, however, we calculate the ratio of fractions
$\Gamma_i / \Gamma_{\textrm{total}}$ for these two decay modes using the numbers and correlation coefficients
published by the Particle Data Group \cite{pdg}, we arrive at
\begin{equation}
r_{3}^{\textrm{(exp)}} = 2.12 \pm 0.19
\end{equation}
by means of standard error propagation.
Such a large branching ratio would indicate significant isospin-violating contributions
in the amplitude. But the experimental uncertainties are sizable
and should be reduced by the upcoming experiments with WASA at COSY \cite{WASA} and at MAMI-C \cite{Nef}.

To summarize the results for the hadronic decays of $\eta$ and $\eta'$ in our approach,
we obtain very good agreement with currently available data
on the decay widths and spectral shapes. In fact, we observe four different classes
of fits which describe these data equally well, but differ in their predictions for yet unmeasured
quantities such as the $\eta' \to \pi^+ \pi^- \pi^0$ decay width (for which there exists
only a weak upper limit) and the Dalitz slope parameters of $\eta' \to 3 \pi$.
The results obtained may be tested in future experiments foreseen at WASA@COSY and MAMI-C.
The hadronic decays considered here along with phase shifts in meson-meson
scattering pose therefore tight constraints on the approach and will allow to 
determine the couplings of the effective Lagrangian up to fourth chiral order.
It is important to stress that the values of the parameters obtained from the fit
are in general not the same as in the framework of ChPT which can be traced back to the
absorption of loops into the coefficients and higher order effects not included
in the perturbative framework.

An intriguing feature of the fits is that they accommodate the large negative
slope parameter $g$ of the decay $\eta \to 3 \pi^0$ measured by the 
Crystal Ball Collaboration \cite{Tippens} which could not be met by previous
theoretical investigations \cite{Kambor, BB1}. 
This value must, however, be confronted with the more recent
but yet preliminary $g$ value of the KLOE Collaboration \cite{KLOE}.
If we replace the PDG data by the KLOE Dalitz parameters of both the charged and neutral
$\eta \to 3 \pi$ decay, we do not achieve
a good overall fit. It appears that the slope parameters of both $\eta \to 3 \pi$ decays
cannot be fitted simultaneously. In addition, fitting to the KLOE data
destroys the agreement with the experimental branching ratio
of both decays which is known to high precision. To this end, we have illustrated that
utilizing the $\Delta I =1$ selection rule which relates both decays and taking 
the KLOE parametrization of the charged decay as input
leads in a model-independent way to a $g$ value not consistent with the KLOE $g$ result.

The importance of the various two-particle channels with different isospin
and angular momentum has been examined as well. For the $\eta \to 3 \pi$ decays we find that the major
contribution is given by $\pi \pi$ rescattering in the $s$-wave $I=0$ channel,
while the $I=1,2$ channels interfere destructively with the former. The $p$-wave
contribution in the charged decay is tiny, since available phase space is small 
and the $C$-odd channels related to the $\rho(770)$ resonance do not occur.
For $\eta' \to \pi^+ \pi^- \pi^0$, on the other hand, phase space is considerably
larger, and the size of the $p$-wave contributions ranges from 10\,\% to 50\,\%
depending on the cluster of fits.

For the decays $\eta' \to \eta \pi \pi$ we find that the
$s$-wave $I=1$ channels dominate for two classes of fits which
would confirm the importance of the nearby $a_0(980)$ resonance as claimed
by previous investigations \cite{Fariborz, BB2, BB1}. But the other two clusters are dominated
by the $I=0$ channels. These two scenarios can be distinguished by their predictions
for the $\eta' \to 3 \pi$ decays. Thus, a precise measurement of $\eta' \to 3 \pi$ decay 
parameters can also help to clarify the importance of $a_0(980)$ or $f_0(980)$ resonance contributions 
to the dominant decay mode of the $\eta'$ into $\eta \pi \pi$.

\begin{ack}
This work was supported in part by DFG, SFB/TR-16 ``Subnuclear Structure of Matter'', and
Forschungszentrum J\"ulich.
\end{ack}

\end{document}